\documentclass[conference]{IEEEtran}
\IEEEoverridecommandlockouts
\usepackage{cite}
\usepackage{amsmath,amssymb,amsfonts}
\usepackage{algorithmic}
\usepackage{graphicx}
\usepackage{tikz}
\usepackage{textcomp}
\usepackage{hyperref}
\usepackage{lipsum}
\usepackage{xcolor}
\def\BibTeX{{\rm B\kern-.05em{\sc i\kern-.025em b}\kern-.08em
    T\kern-.1667em\lower.7ex\hbox{E}\kern-.125emX}}
    
\newcommand\copyrighttext{%
  \footnotesize \textcopyright 2021 IEEE. Personal use of this material is permitted. Permission from IEEE must be obtained for all other uses, in any current or future media, including reprinting/republishing this material for advertising or promotional purposes, creating new collective works, for resale or redistribution to servers or lists, or reuse of any copyrighted component of this work in other works. DOI: 10.1109/I2MTC50364.2021.9460087}
\newcommand\copyrightnotice{%
\begin{tikzpicture}[remember picture,overlay]
\node[anchor=south,yshift=10pt] at (current page.south) {\fbox{\parbox{\dimexpr\textwidth-\fboxsep-\fboxrule\relax}{\copyrighttext}}};
\end{tikzpicture}%
}
\begin{document}

\title{Fast 3-dimensional estimation of the Foveal Avascular Zone from OCTA \\

\thanks{NISA Study was funded by grants from the College of Optometrists. Diabetes UK, Macular Society, Guidedogs for the Blind, Thomas Pocklington Trust and the Belfast Association for the Blind. For NICOLA funding see\cite{b15}. }
}

\author{\IEEEauthorblockN{1\textsuperscript{st} Giovanni Ometto}
\IEEEauthorblockA{\textit{Division of Optometry \& Visual Sciences} \\
\textit{City,  University  of  London}\\
London, United kingdom\\
giovanni.ometto@city.ac.uk}
\and
\IEEEauthorblockN{2\textsuperscript{nd} Giovanni Montesano}
\IEEEauthorblockA{\textit{Division of Optometry \& Visual Sciences} \\
\textit{City,  University  of  London}\\
London, United kingdom\\
giovmontesano@gmail.com}
\and
\IEEEauthorblockN{3\textsuperscript{rd} Usha Chakravarthy}
\IEEEauthorblockA{\textit{Centre for Public Health} \\
\textit{Queen's University Belfast}\\
Belfast, United kingdom\\
U.Chakravarthy@qub.ac.uk}
\and
\IEEEauthorblockN{4\textsuperscript{th} Frank Kee}
\IEEEauthorblockA{\textit{Centre for Public Health} \\
\textit{Queen's University Belfast}\\
Belfast, United kingdom\\
F.Kee@qub.ac.uk}
\and
\IEEEauthorblockN{5\textsuperscript{th} Ruth E. Hogg}
\IEEEauthorblockA{\textit{Centre for Public Health} \\
\textit{Queen's University Belfast}\\
Belfast, United kingdom\\
r.e.hogg@qub.ac.uk}
\and
\IEEEauthorblockN{6\textsuperscript{th} David P. Crabb}
\IEEEauthorblockA{\textit{Division of Optometry \& Visual Sciences} \\
\textit{City, University of London}\\
London, United kingdom\\
David.Crabb.1@city.ac.uk}
}

\maketitle
\copyrightnotice

\begin{abstract}
The area of the foveal avascular zone (FAZ) from en face images of optical coherence tomography angiography (OCTA) is one of the most common measurement based on this technology. However, its use in clinic is limited by the high variation of the FAZ area across normal subjects, while the calculation of the volumetric measurement of the FAZ is limited by the high noise that characterizes OCTA scans. We designed an algorithm that exploits the higher signal-to-noise ratio of en face images to efficiently identify the capillary network of the inner retina in 3-dimensions (3D), under the assumption that the capillaries in separate plexuses do not overlap. The network is then processed with morphological operations to identify the 3D FAZ within the bounding segmentations of the inner retina. The FAZ volume and area in different plexuses were calculated for a dataset of 430 eyes. Then, the measurements were analyzed using linear mixed effect models to identify differences between three groups of eyes: healthy, diabetic without diabetic retinopathy (DR) and diabetic with DR. Results showed significant differences in the FAZ volume between the different groups but not in the area measurements. These results suggest that volumetric FAZ could be a better diagnostic detector than the planar FAZ. The efficient methodology that we introduced could allow the fast calculation of the FAZ volume in clinics, as well as providing the 3D segmentation of the capillary network of the inner retina.
\end{abstract}


\section{Introduction}
Optical coherence tomography angiography (OCTA) is a recent technology that allows non-invasive imaging of retinal microvasculature. In short, OCTA measures the decorrelation signal between successive OCT scans (b-scans) to highlight areas where motion is present. This process produces depth-resolved maps of the flow of erythrocytes inside blood vessels \cite{b1,b2}.

OCTA has transformed the evaluation of diseases that affect retinal microvasculature by allowing the development of objective, quantitative biomarkers. The most used biomarkers that can be obtained with this technology include vessel density, fractal dimensions and length; intracapillary spaces; avascular areas; and detection and measurement of the foveal avascular zone (FAZ) area. The measurement of the FAZ for the qualitative assessment of nonperfusion associated with diabetic retinopathy (DR) is one of the most common uses of OCTA \cite{b2}. In fact, imaging of the FAZ with OCTA is more suitable than fluorescein angiography (FA) and indocyanine green angiography (IGA) due to the lack of leakage and clearer capillary detailing, without the need of dye injections \cite{b2,b3,b4,b5}.

FAZ is commonly identified through the analysis of 2-dimensional (2D), maximum projections of OCTA slabs. It has been demonstrated that these projections improve contrast and signal-to-noise ratio (SNR) of OCTA \cite{b6}. En face views are calculated using the segmentation of anatomically relevant slabs after the removal of projection artifacts \cite{b3}, producing angiograms in a format similar to FA / IGA. In literature, FAZ is often calculated from en face projections of the superficial vascular complex (SVC) or deep capillary plexus (DCP) though restricting the measurement solely to these slabs is not well justified. Published studies have revealed a large variation in 2D FAZ measurements, even among healthy subjects \cite{b3, b5}. These results have driven recent efforts to develop and identify better detectors of foveal ischemia using OCTA \cite{b4, b7, b8}. 
However, it is recognized that there has been little effort in the development of 3D algorithms that can clearly and reliably identify the capillary network and, consequently, the volumetric FAZ. It is also recognized that the problem lies in the complexity of the analysis of the volumetric OCTA signal, characterized by a considerable amount of noise \cite{b2}.

With this work we introduce a novel algorithm for the fast estimation of the FAZ in 3D. We achieve this by exploiting the morphology of the inner retinal vasculature, allowing the introduction of a key assumption that simplifies the problem. We can then robustly and efficiently identify the 3D network by an extensive use of morphological operations based on the 2D maximum projections, where contrast and SNR is higher, and identify a volumetric FAZ.
Finally, we used the proposed algorithm to evaluate differences between the 3D and 2D FAZ measurements in in a large cohort of healthy subjects, and diabetics with and without DR.

\section{Related Work}

Several works have been published around the measurement of the FAZ in 2D from OCTA, since the recent introduction of this technology in clinics. A detailed review can be found in \cite{b2, b3}. Efforts to improve and automatize the measurement are ongoing, with new publications exploiting the latest technologies \cite{b8, b9, b10, b11, b12, b13}. However, only a few of these publications are based on 3D measurements from OCTA \cite{b8, b13}, and no attempt to segment the FAZ in 3D has been made yet.

\section{Materials and Methods}

\subsection{Dataset}
The Northern Ireland Sensory Ageing Study (NISA) \cite{b14}, a follow-up to the Northern Ireland Cohort for the Longitudinal Study of Ageing (NICOLA) \cite{b15}, is a prospective, population-based study of early imaging and functional biomarkers of DR and age-related macular degeneration (AMD). The sample of participants was  enriched with patients with diabetes (type I or type II) recruited concurrently from the Belfast Trust Diabetic Retinopathy Hospital Clinics at QUB were later added to the dataset. A number of functional test and measurements were acquired during the visits: Best Corrected Visual Acuity; Contrast Sensitivity; mesopic microperimetry and Frequency Doubling Technology perimetry; an SD-OCT scan (Spectralis SD-OCT, Heidelberg Engineering, Heidelberg, Germany); fundus colour picture (CX-1 Digital Fundus Camera, Canon U.S.A., Inc, Tokyo, Japan); and axial length (AL, Lenstar LS 900 Biometer, Haag-Streit AG, Switzerland). An OCTA volume scan (Spectralis SD-OCT) was also acquired for a subgroup of the participants. Dilation was induced pharmacologically with tropicamide 1\% before any imaging procedure. 

\subsubsection{OCTA data}
A 10×10 degrees volume scan pattern centred on the fovea was used for the OCTA imaging acquisition with the Spectralis SD-OCT. Each volume included 512 successive, horizontal b-scans of 496×512 pixels of resolution (axial × horizontal) for a total volume of 512×512×496 pixels. Therefore, the 512×512 pixels, planar resolution of these volumes was isometric (with varying resolution across eyes, depending on the AL), while the axial resolution was 3.87 $\mu$m (instrument resolution, provided by the manufacturer). All OCTA scans were processed by the Projection Artefacts Removal (PAR) tool (Heidelberg Engineering, Heidelberg, Germany) before export.
The OCTA segmentations of the Inner Limiting Membrane (ILM), Inner Plexiform Layer (IPL), Outer Plexiform Layer (OPL) and Retinal Pigment Epithelium (RPE), as well as the OCTA en face images, were provided by the manufacturer of the device and were obtained from .vol files generated with a version of the Heyex software (Heidelberg Engineering, Heidelberg, Germany) enabled for RAW data export. 

\subsubsection{Clinical classification}
Two clinical graders examined all fundus photographs to identify signs of pathologies. If participants were not recruited from the DR clinic, self-reported diagnosis of diabetes was used for their classification as diabetic. Participants aged 50 years or over and all subjects classified as diabetic were invited to take a blood test to measure concentration of plasma glycated haemoglobin (HbA1C). A HbA1C concentration of over 6.5\% was used as to classify the participants as diabetic when another diagnosis was not available. Blood test refusal did not prevent inclusion.

\subsubsection{OCTA dataset selection}
For this work, only OCTA volume scans that were judged of sufficient quality by a manual classification were considered. Out of 686 scans available, 483 were manually identified. Of these, 41 scans were excluded, as showing signs of intermediate or advanced AMD. This resulted in a dataset of 442 scans from eyes of healthy participants and diabetics with or without DR. A further selection excluded 12 OCTA scans from eyes with macular oedema and/or segmentation errors that could have affected the measurement. The final dataset included 430 OCTA scans, with 317 eyes from 189 healthy controls and 113 eyes from 82 people with diabetes. Of these, 20 eyes had DR. Five patients had DR in one eye and no DR in the fellow eye. Descriptive statistics of the selected sample are reported in (Table ~\ref{table1}).

\subsection{Image processing}
\begin{figure}[tb]
\centerline{\includegraphics{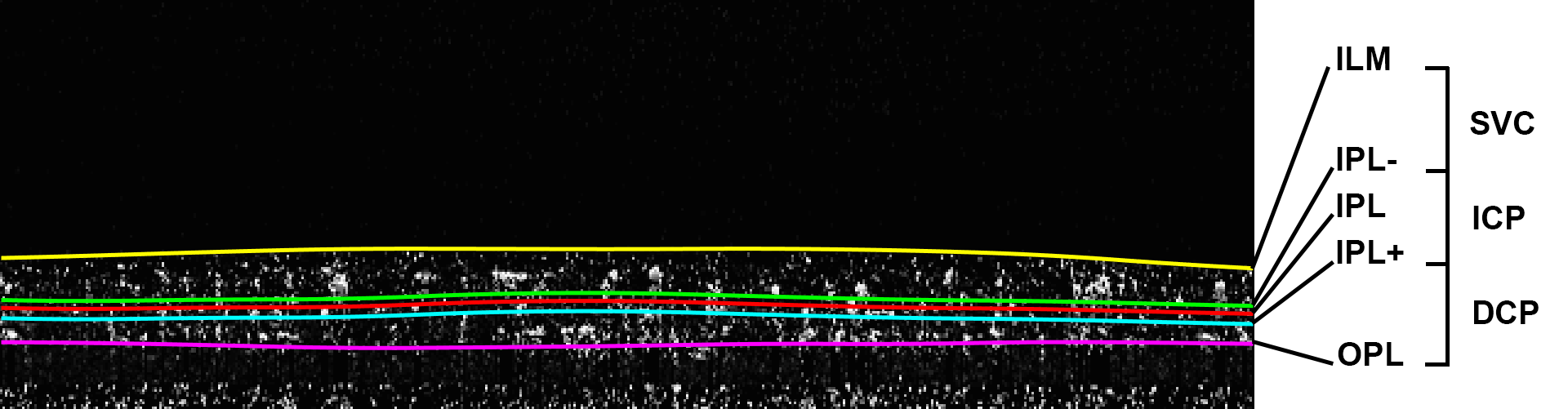}}
\caption{Inner retina layers in an OCTA b-scan and the three plexuses: superficial vascular complex (SVC); intermediate capillary plexus (ICP); and deep capillary plexus (DCP). ILM = inner limiting membrane; IPL = inner plexiform layer; OPL = outer plexiform layer.}
\label{segmentations}
\end{figure}

The proposed algorithm identifies the vascular network of the inner retina as the combination of three vascular networks, each calculated independently. These networks are in three different plexuses: superficial vascular complex (SVC); intermediate capillary plexus (ICP); and deep capillary plexus (DCP)\cite{b16}. Plexuses are delimited by segmentation of retinal layers or from their derivations: SVC located between the ILM and a posterior offset of -17 $\mu$m to the lower boundary of the IPL (IPL–), ICP between IPL– and a +22 $\mu$m offset to the lower boundary of the IPL (IPL+), DCP between IPL+ and OPL (Fig.~\ref{segmentations}). These boundaries are used for the generation of the maximum projection en face images from the OCTA volume. Within each plexus, the overwhelming majority of vessels of the same network do not overlap, and this is a key working assumption for the algorithm. This is coherent with the morphology of parafoveal capillary layout, except for inter-plexuses capillary connections. However, these connections are not depicted in OCTA images since they are almost invariably masked by signal from overlying vessels. Crucially, such an assumption imposes a one-to-one association between a vessel-location on the planar, en face image and the location on the 3-dimensional OCTA scan within the boundary of its relative plexus. Having defined this association, vessels of one network could first be identified on the en face image, a relatively simple task given the higher contrast and SNR. Then, the axial position of each identified vessel-location could be obtained from the OCTA volume for a full 3-dimensional representation. Finally, the combination of the three networks, with the ILM and OPL segmentations, could be used to define the 3D FAZ. All image processing operations were performed using Matlab R2020a software (MathWorks, Natick, MA, United States) with the Image Processing Toolbox.

\begin{table}[tb]
\caption{Descriptive statistics of the selected sample}
\begin{center}
\begin{tabular}{c|c|c|c|}
\cline{2-4}
                                                                                            & \textbf{\begin{tabular}[c]{@{}c@{}}Healthy \\ (N = 317)\end{tabular}} & \textbf{\begin{tabular}[c]{@{}c@{}}Diabetes w/o \\ DR$^{\mathrm{d}}$ (N = 93)\end{tabular}} & \textbf{\begin{tabular}[c]{@{}c@{}}Diabetes w/ \\ DR$^{\mathrm{d}}$ (N = 20)\end{tabular}} \\ \hline
\multicolumn{1}{|c|}{\textbf{\begin{tabular}[c]{@{}c@{}}Age$^{\mathrm{a}}$\\ (years)\end{tabular}}}        & \begin{tabular}[c]{@{}c@{}}65 \\ {[}60, 73{]}\end{tabular}            & \begin{tabular}[c]{@{}c@{}}69 \\ {[}63, 74{]}\end{tabular}                   & \begin{tabular}[c]{@{}c@{}}67 \\ {[}60, 71{]}\end{tabular}                  \\ \hline
\multicolumn{1}{|c|}{\textbf{\begin{tabular}[c]{@{}c@{}}Sex$^{\mathrm{a}}$ \\ (Male:Female)\end{tabular}}} & 81:107                                                                & 8:54                                                                         & 8:11                                                                        \\ \hline
\multicolumn{1}{|c|}{\textbf{\begin{tabular}[c]{@{}c@{}}HbA1C$^{\mathrm{a,b}}$ \\ (\%)\end{tabular}}}        & \begin{tabular}[c]{@{}c@{}}5.5 \\ {[}5.29, 5.73{]}\end{tabular}       & \begin{tabular}[c]{@{}c@{}}6.78 \\ {[}6.24, 8.02{]}\end{tabular}             & \begin{tabular}[c]{@{}c@{}}8.54 \\ {[}6.86, 8.72{]}\end{tabular}            \\ \hline
\multicolumn{1}{|c|}{\textbf{\begin{tabular}[c]{@{}c@{}}BCVA$^{\mathrm{a,c}}$ \\ (Letters)\end{tabular}}}    & \begin{tabular}[c]{@{}c@{}}85 \\ {[}80, 89{]}\end{tabular}            & \begin{tabular}[c]{@{}c@{}}87 \\ {[}82, 90{]}\end{tabular}                   & \begin{tabular}[c]{@{}c@{}}83 \\ {[}82, 87{]}\end{tabular}                  \\ \hline
\multicolumn{1}{|c|}{\textbf{\begin{tabular}[c]{@{}c@{}}Axial length$^{\mathrm{a}}$\\ (mm)\end{tabular}}}  & \begin{tabular}[c]{@{}c@{}}23.46 \\ {[}22.95, 23.9{]}\end{tabular}    & \begin{tabular}[c]{@{}c@{}}23.32 \\ {[}22.8, 24.36{]}\end{tabular}           & \begin{tabular}[c]{@{}c@{}}23.65 \\ {[}23.09, 24{]}\end{tabular}            \\ \hline
\multicolumn{4}{l}{$^{\mathrm{a}}$Median [interquartile range].}\\
\multicolumn{4}{l}{$^{\mathrm{b}}$HbA1C  = glycated hemoglobin.}\\
\multicolumn{4}{l}{$^{\mathrm{c}}$BCVA = Best Corrected Visual Acuity.}\\
\multicolumn{4}{l}{$^{\mathrm{d}}$DR = Diabetic Retinopathy.}
\end{tabular}
\label{table1}
\end{center}
\end{table}

\subsubsection{OCTA and segmentation pre-processing}
OCTA volumes were resized along the axial dimension to match the planar resolution and make voxels isometric. The planar resolution in $\mu$ m ($res_{plane}$) was calculated correcting for the axial length with the procedure described in \cite{b17}. The calculated $res_{plane}$ and the known axial resolution of the instrument (3.87 $\mu$m) provided the rescaling ratio $R$ for the axial dimension: $R =$ 3.87 $\mu$m / $res_{plane}$. Segmentations were also rescaled accordingly and were regularized by the identification of outliers through a median moving window of 15×15 pixels in size and by their replacement with values obtained with cubic spline interpolation. Each column of each b-scan was then shifted to move the RPE location to the bottom of the scan, “flattening” the retina on this layer. Finally, the volume was filtered using a 3D Gaussian filter ($\sigma = 3$) to reduce noise and to incorporate the information of adjacent scans into each b-scan. 

\subsubsection{2D network segmentation}
\begin{figure*}[tbph]
\centerline{\includegraphics[width=15.5cm]{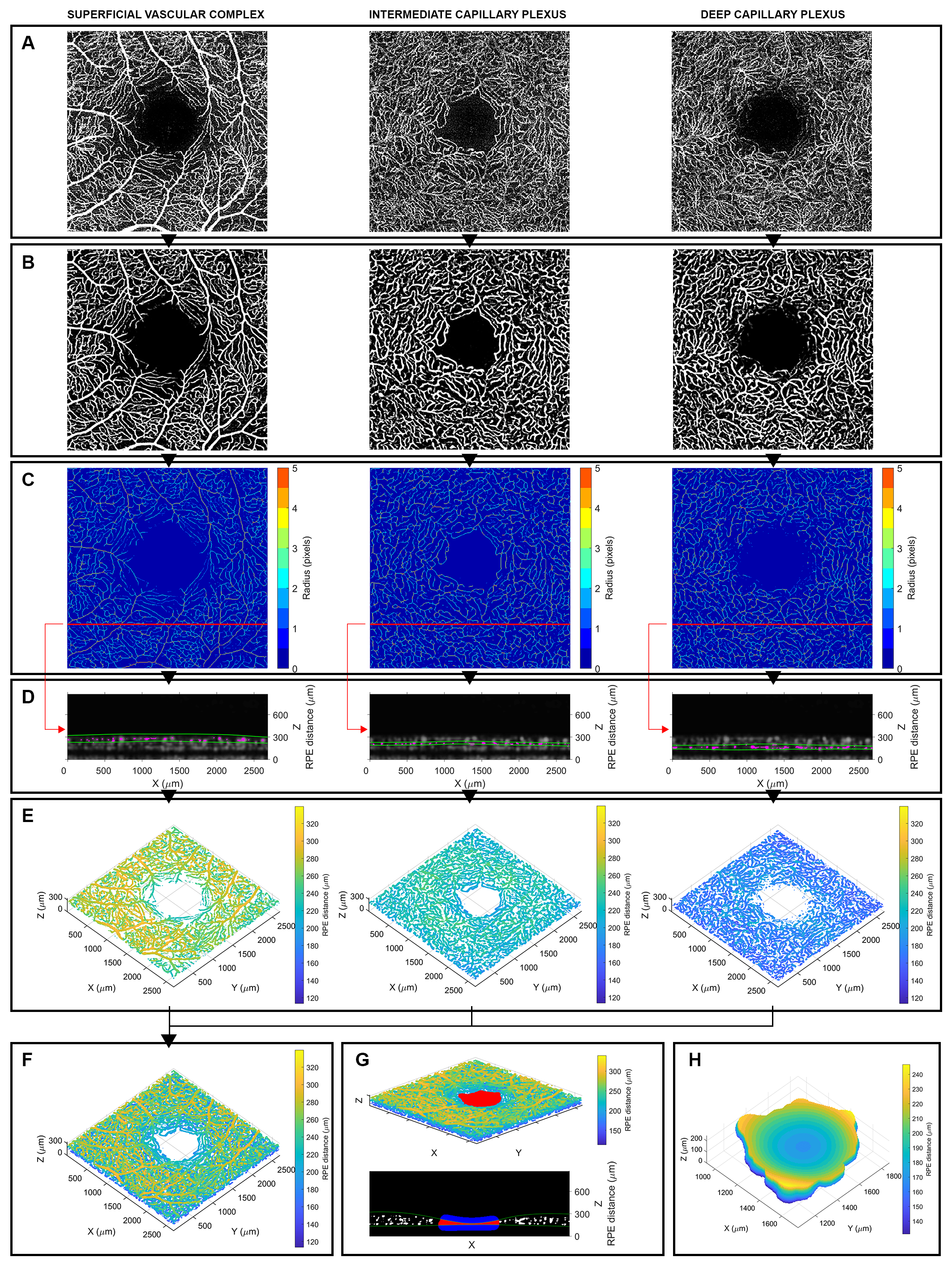}}
\caption{A visualization of the processing operations in a random OCTA of the dataset. A) En face maximum projection images. B) Vessel-enhanced images. C) Vasculature skeletonization with radius map. D) A random pre-processed b-scan showing the axial location of the vessels (in magenta) for each location of the skeletonized map, within the respective bounding segmentations (in green). E) Reconstructed 3-dimensional (3D) capillary network for the individual plexuses. F) Reconstructed cumulative capillary network of the inner retina. G) Calculated 3D Foveal Avascular Zone (FAZ, in red) depicted in the 3D capillary network (top) and in the central, 256\textsuperscript{th} b-scan along the Y dimension (bottom). In the b-scan, the blue and red areas represent a section of the dilated FAZ volume, that is then restricted to the volume between the segmentations of the inner limiting membrane and outer plexiform layer (in green). H) The calculated FAZ 3D volume.}
\label{diagram}
\end{figure*}

The same process for the identification of each of the three capillary networks within their relative plexus (SVC, ICP, DCP) was performed independently (Fig.~\ref{diagram}A-E). The process started from the en face, maximum-intensity-projection provided by the manufacturer and obtained from the OCTA, and the segmentations of the respective bounding layers: ILM to IPL– for the superior; IPL– to IPL+ for the intermediate; and IPL+ to OPL for the deep plexus (Fig.~\ref{diagram}A). The three, en face images were processed with a multi-scale vessel enhancing algorithm \cite{b18} to remove noise and enhance the signal from vessel-like structures (Fig.~\ref{diagram}B). Parameters used in the process were identified experimentally, accounting for the morphological characteristics of the microvasculature depicted in these images: $ScaleRange = [2;3]$; $ScaleRatio = 1$; $BetaOne = 0.6$; $BetaTwo = 22$. Then, vessel enhanced images were binarized using a global histogram threshold calculated with Otsu's method \cite{b19}. Binary images were skeletonized, and, for each logical $true$ pixel of the skeleton ($sk(x,y)$), the distance transform was calculated with a two-pass, sequential scanning algorithm \cite{b20} (Fig.~\ref{diagram}C). This process was used to obtain an estimate of radiuses of vessels at each location of the skeleton ( $r_{sk}$ = $radius(sk(x,y))$ ), when the skeleton is interpreted as the centerline of the vessel and assuming that vessels have a tubular shape with circular section. For the regularization of inconsistent, adjacent radius sizes, the distance transform was filtered with a Gaussian filter ($\sigma = 1$), then discretized to integers with rounding. As a result of this process, the superficial, intermediate and deep capillary networks were fully described by their centerline locations ($sk_{sup}(x,y)$, $sk_{int}(x,y)$, $sk_{deep}(x,y)$) and radiuses ($r_{sk_{sup}}$, $r_{sk_{int}}$, $r_{sk_{deep}}$) in 2-dimensions, with a resolution equal to the planar resolution of the scans. 

\subsubsection{3D network segmentation} 
The 3-dimensional reconstruction was based on the search of the coordinate z in the third, axial dimension for each 2D skeleton ($sk_{sup}$, $sk_{int}$, $sk_{deep}$). Under the aforementioned assumption derived from the morphology, no more than one vessel could be located in a single plexus for a single location of the centerline $sk(x,y)$. Therefore, the search of the axial coordinate was performed on the pre-processed OCTA volume ($OCTA(x,y,z)$), using the segmentations to restrict the search field on the axial dimension (Fig.~\ref{diagram}D). For each location $(x,y)$ of $sk_{sup}$, $sk_{int}$ and $sk_{deep}$, the index of the maximum value $z_{sk}$, in the corresponding vertical column of the OCTA volume and between the boundaries of the plexus, was defined by:

\begin{equation}
\begin{aligned}[b]
z_{sk_{sup}} = {}   & loc( max(OCTA(x,y,z) ) )\\
                    & \forall (x,y) \in sk_{sup}; \mathrm{IPL-} < z_{sk_{sup}} < \mathrm{ILM},\label{eq1}
\end{aligned}
\end{equation}

\begin{equation}
\begin{aligned}[b]
z_{sk_{int}} = {}   & loc( max(OCTA(x,y,z) ) )\\
                    & \forall (x,y) \in sk_{int}; \mathrm{IPL+} < z_{sk_{int}} < \mathrm{IPL-},\label{eq2}
\end{aligned}
\end{equation}

\begin{equation}
\begin{aligned}[b]
z_{sk_{deep}} = {}  & loc( max(OCTA(x,y,z) ) )\\
                    & \forall (x,y) \in sk_{deep}; \mathrm{OPL} < z_{sk_{deep}} < \mathrm{IPL+},\label{eq3}
\end{aligned}
\end{equation}
where $loc$ is a function that returns the index of an element in an array.
In the detection of the z-coordinate, the pre-processing of the OCTA volume with a 3D Gaussian filter helped reducing the detection of spurious, local-maxima values. Of notice, all scans are already pre-processed by a proprietary algorithm to remove projection artefacts upon extraction. Then, each resulting 3D skeletonization, $sk_{sup}(x,y,z)$, $sk_{int}(x,y,z)$ and $sk_{deep}(x,y,z)$ was separated into $n$ subsets based on the radius-values measured in the 2D maps: $sk(x,y,z,r_{n})$, $n = 1,2,...,max(r_{sk})$. Each of these $sk(x,y,z,r_{n})$ was convoluted with a spherical structuring element of radius $r_{n}$ to represent the 3D size of vessel segments of the same size. The $n$, logical, 3-dimensional sets of coordinates obtained with this operation were combined back together using the logical OR operation to represent the superior, intermediate and deep capillary networks (Fig.~\ref{diagram}E). Finally, these three networks were combined with another logical OR to represent the full capillary network of the inner retina (Fig.~\ref{diagram}F).

\subsubsection{Segmentation and measurement of 2D FAZ} 
The binarized images obtained after the vessel-enhancement procedure were used for the identification of the FAZ in 2-dimensions. Connected components smaller than 5 pixels were removed from the binary maps to reduce false, spurious detection of the vasculature. Then, the map was dilated with a morphological structuring disk of radius 15 pixels. The radius was determined experimentally to fill the gaps between the vascular structures. Then, the largest connected component of the inverted, resulting map was taken as the eroded FAZ. A dilation operation with the same structuring disk used in the dilation of the vasculature produced the 2D FAZ. The number of pixels of the FAZ multiplied by the area of a pixel ($res_{plane}^{2}$) provided the area of the FAZ in mm$^{2}$.

\subsubsection{Segmentation and measurement of 3D FAZ} 
The procedure in 3D was equivalent to that described for the FAZ in the 2D domain, except for the use of the calculated 3D network representation in place of binary images, for the replacement of the disk with a spherical structural element (of equal radius), and for the axial restriction of the volume between the segmentations of the ILM and OPL (Fig.~\ref{diagram}G). Finally, the volume of the FAZ in mm$^{3}$ was estimated multiplying the number of voxels by the known size of a voxel ($res_{plane}^{3}$) (Fig. 3H). 

\subsection{Evaluation}
The volume of the FAZ in 3D and the areas of the FAZ in 2D from the three vascular plexuses were calculated with the proposed algorithm for all the OCTA scans in the dataset. Results were displayed superimposing the 2D FAZ on the vessel-enhanced images and as a point cloud plot for the 3D FAZ (as shown in Fig.~\ref{diagram}G-H) to be manually inspected for errors by a clinical grader [G. Montesano]. Then, the measurements were analyzed to identify differences between the three groups: healthy, diabetic without DR and diabetic with DR. All statistical analyses were performed in R (R Foundation for Statistical Computing, Vienna, Austria) using linear mixed effect models. Random intercepts were used to account for correlations between the two eyes from the same subject. The statistical significance level was set at 0.05. The results are reported as mean estimate [95\% Confidence Intervals].

\section{Results}
The algorithm successfully calculated all FAZ in the dataset. The inspection of the results by a clinical grader found no errors (100\% success rate). Mean (standard deviation; SD) processing time was 38.7 (2.6) seconds per OCTA volume, running on a laptop computer with an Intel (Santa Clara, CA, US) Core i7-1065G7 CPU @ 1.30 GHz and 16 MB of RAM memory. 

\begin{figure}[tbph]
\centerline{\includegraphics{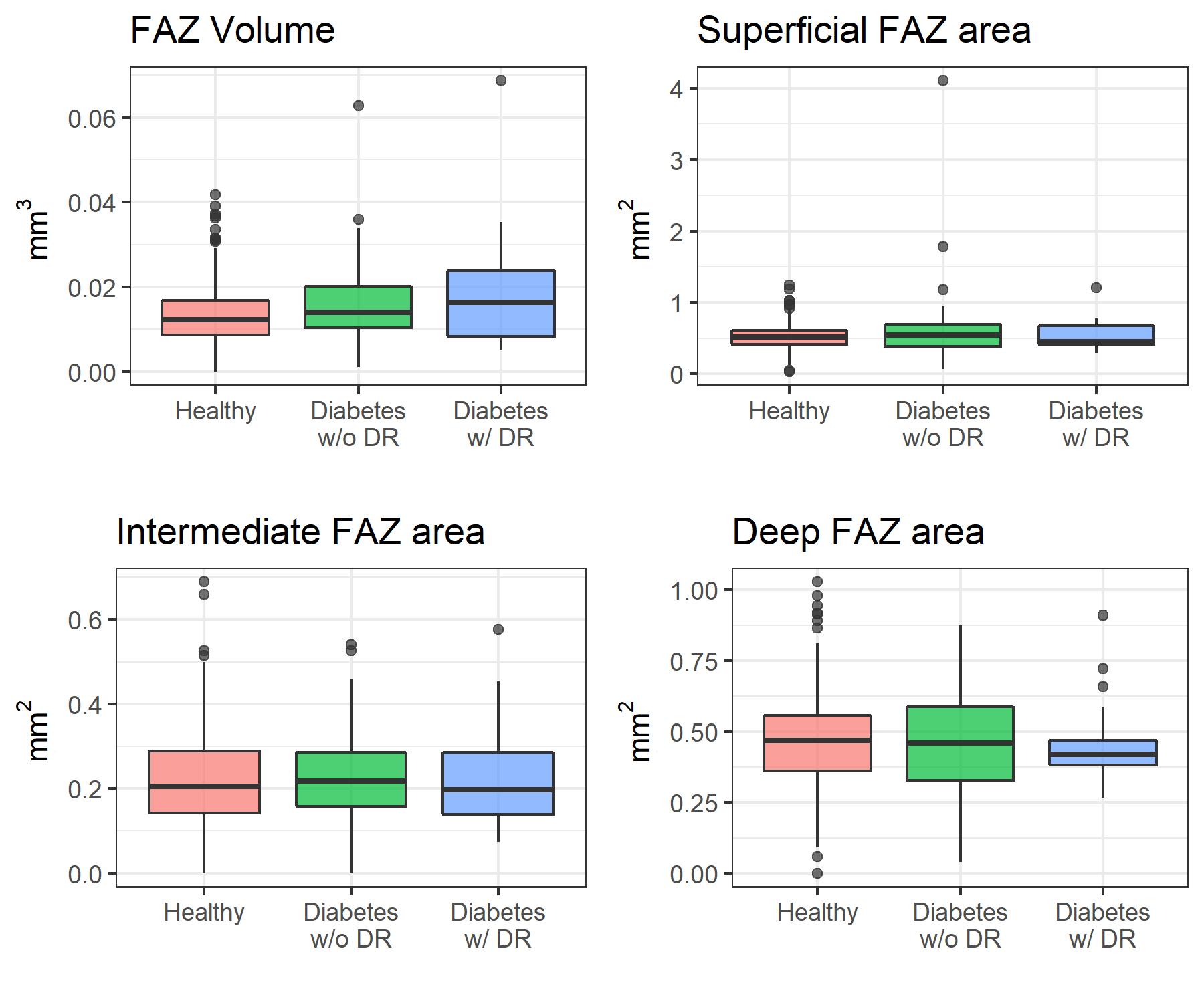}}
\caption{Boxplot of the distribution of the calculated FAZ volume and area for the three groups in the dataset. Boxes extends from the 25\textsuperscript{th} to 75\textsuperscript{th} percentile (inter-quartile range, IQR), with the mid-line representing the median. Whiskers extends to a maximum of 1.5 * IQR from the edges of the box. Dots represent outliers. FAZ = Foveal Avascular Zone. DR = Diabetic Retinopathy}
\label{boxplot}
\end{figure}

The results of the linear mixed effect model are reported in Table~\ref{table2} using the mean estimate [95\% Confidence Intervals]. 
There was no statistically significant difference between the different groups for the FAZ area calculated from different vascular plexuses. On the contrary, for both groups of diabetic eyes, there was a significant difference in the mean FAZ volume compared to the healthy group (Fig.~\ref{boxplot}). 

\begin{table}[tbph]
\caption{Measurement difference between groups}
\begin{center}
\begin{tabular}{c|c|c|c|c|c|}
\cline{2-6}
                                                                                                       & \textbf{Healthy}                                                     & \textbf{\begin{tabular}[c]{@{}c@{}}Diabetes \\ w/o DR$^{\mathrm{b}}$\end{tabular}}   & \textbf{\begin{tabular}[c]{@{}c@{}}p\\ value\end{tabular}} & \textbf{\begin{tabular}[c]{@{}c@{}}Diabetes \\ w/ DR$^{\mathrm{b}}$\end{tabular}}    & \textbf{\begin{tabular}[c]{@{}c@{}}p\\ value\end{tabular}} \\ \hline
\multicolumn{1}{|c|}{\textbf{\begin{tabular}[c]{@{}c@{}}FAZ\\ Volume$^{\mathrm{a}}$\\ (mm$^{3}$)\end{tabular}}}            & \begin{tabular}[c]{@{}c@{}}0.013\\ {[}0.012,\\ 0.015{]}\end{tabular} & \begin{tabular}[c]{@{}c@{}}0.016 \\ {[}0.014,\\ 0.018{]}\end{tabular} & 0.032$^{\mathrm{*}}$                                                      & \begin{tabular}[c]{@{}c@{}}0.018\\ {[}0.014,\\ 0.021{]}\end{tabular}  & 0.013$^{\mathrm{*}}$                                                      \\ \hline
\multicolumn{1}{|c|}{\textbf{\begin{tabular}[c]{@{}c@{}}Superficial\\ FAZ area$^{\mathrm{a}}$ \\ (mm$^{2}$)\end{tabular}}} & \begin{tabular}[c]{@{}c@{}}0.525\\ {[}0.492,\\ 0.557{]}\end{tabular} & \begin{tabular}[c]{@{}c@{}}0.580\\ {[}0.522,\\ 0.637{]}\end{tabular}   & 0.102                                                      & \begin{tabular}[c]{@{}c@{}}0.537\\ {[}0.438, \\ 0.636{]}\end{tabular} & 0.817                                                      \\ \hline
\multicolumn{1}{|c|}{\textbf{\begin{tabular}[c]{@{}c@{}}Intermediate\\ FAZ area$^{\mathrm{a}}$\\ (mm$^{2}$)\end{tabular}}} & \begin{tabular}[c]{@{}c@{}}0.216\\ {[}0.200,\\ 0.231{]}\end{tabular}   & \begin{tabular}[c]{@{}c@{}}0.224\\ {[}0.198,\\ 0.250{]}\end{tabular}   & 0.585                                                      & \begin{tabular}[c]{@{}c@{}}0.218\\ {[}0.180,\\ 0.255{]}\end{tabular}   & 0.922                                                      \\ \hline
\multicolumn{1}{|c|}{\textbf{\begin{tabular}[c]{@{}c@{}}Deep\\ FAZ area$^{\mathrm{a}}$\\ (mm$^{2}$)\end{tabular}}}         & \begin{tabular}[c]{@{}c@{}}0.465\\ {[}0.443,\\ 0.487{]}\end{tabular} & \begin{tabular}[c]{@{}c@{}}0.459\\ {[}0.420,\\ 0.497{]}\end{tabular}   & 0.774                                                      & \begin{tabular}[c]{@{}c@{}}0.453\\ {[}0.394,\\ 0.513{]}\end{tabular}  & 0.708                                                      \\ \hline
\multicolumn{6}{l}{$^{\mathrm{a}}$Estimates [95\% Confidence Intervals]. FAZ = Foveal Avascular Zone}\\
\multicolumn{6}{l}{$^{\mathrm{b}}$DR = Diabetic Retinopathy.}\\
\multicolumn{6}{l}{$^{\mathrm{*}}$Statistically significant (significance level = 0.05).}
\end{tabular}
\label{table2}
\end{center}
\end{table}

\section{Discussion}
The fast processing time and success rate of this novel algorithm suggest it is possible to reliably and efficiently segment the 3D FAZ from OCTA. This was likely achieved thanks to the assumption that the capillary vasculature of a single plexus does not overlap, allowing for an important part of the image processing operations to take place in 2-dimensions. The benefits of this were dual: on one hand, the analysis of 2D images helped overcome the problem of low SNR in OCTA volumetric scans; on the other hand, the proposed 2D analysis is fast. Moreover, the discretized, logical representation of the 2D vasculature that we propose makes it possible to extensively use efficient, morphological operations in 3D. Finally, the algorithm provided the full segmentation of the capillary network of the inner retina as a by-product of the process.
Results from the statistical analysis showed that the FAZ volume was significantly different between groups. This was not true for FAZ areas, suggesting that the volume is preferable than the conventional area as a diagnostic measure. Of note, results still show large overlaps between groups. Therefore, the FAZ volume should still be interpreted by physicians in conjunction with other clinical evidence to support diagnostic and treatment decisions. However, an improved characterization of the FAZ can justify and promote further research, not only in diabetes, where it is currently predominately used, but potentially in other diseases as well.
The proposed algorithm has some limitations. First, the algorithm requires the segmentations and the artifact-corrected, en face images as well as the OCTA volume. While these are often offered by the manufacturers for visualization, their export and manipulation are not as widely available. However, if the proposed methodology was implemented within the processing software of acquisition devices, this requirement would not represent a limitation. Second, the algorithm was used to process only manually selected, good-quality scans, using correct segmentations and in absence of edema or macular degeneration. The lack of any of these conditions could lead to inaccurate measurements, either because the input data is inaccurate or because the definition of the FAZ is invalid under these pathological circumstances. Third, some parameters in the algorithm are experimentally determined and the variation of these parameters could change the measurements. However, the use of a fixed, pre-selected parameter over another one could only lead to global changes in the statistical analysis.
The statistical analysis is also limited by unbalanced observation in the three groups. The number of eyes of diabetic patients, and particularly those with presence of DR, is much smaller than healthy eyes and this is reflected in the variance of the 3D FAZ measurements (Fig.~\ref{boxplot}). While the availability of a dataset with a greater number of diabetic eyes would improve the estimate of differences between groups, statistically significant differences of the 3D FAZ were observed, and this highlights the importance of the proposed, volumetric measurement.

\section{Conclusions}
We introduced a novel, fast algorithm for the calculation and measurement of the volumetric Foveal Avascular Zone based on OCTA volume scans. Statistical analysis of measures from this technique on scans from a very large cohort of eyes showed significant differences for the selected groups, indicating potential clinical usefulness. The analysis did not identify differences in the mean area of the FAZ. The efficient methodology that we introduced could offer real-time, 3D measurements of the FAZ for the clinical use, while also providing the 3D segmentation of the capillary network of the inner retina.

\end{document}